\documentclass[epj]{webofc}
\usepackage[varg]{txfonts}   % Web of Conferences font
\wocname{EPJ Web of Conferences}
\woctitle{INPC 2013}
\begin{document}
\title{Three-nucleon forces in exotic open-shell isotopes}
%
% subtitle is optional

\author{V. Som{\`a}\inst{1,2}\fnsep\thanks{\email{vittorio.soma@physik.tu-darmstadt.de}} \and
        C. Barbieri\inst{3} \and
        A. Cipollone\inst{3} \and
        T. Duguet\inst{4,5} \and
        P. Navr\'atil\inst{6}
}

\institute{Institut f{\"u}r Kernphysik, Technische Universit{\"a}t Darmstadt, 64289 Darmstadt, Germany
\and
           EMMI, GSI Helmholtzzentrum f\"ur Schwerionenforschung GmbH, 64291 Darmstadt, Germany
\and
           Department of Physics, University of Surrey, 
Guildford, Surrey GU2 7XH, United Kingdom
\and
	CEA-Saclay, IRFU/Service de Physique Nucl\'eaire, F-91191 Gif-sur-Yvette, France
\and
	NSCL and Department of Physics and Astronomy, Michigan State University, East Lansing, MI 48824, USA
\and
	TRIUMF, 4004 Wesbrook Mall, Vancouver, British Columbia, V6T 2A3, Canada
          }

\abstract{
Advances in the self-consistent Green's function approach to finite nuclei are discussed, including the implementation of three-nucleon forces and the extension to the Gorkov formalism. We report results on binding energies in the nitrogen and fluorine isotopic chains, as well as spectral functions of $^{22}$O. The application to medium-mass open-shell systems is illustrated by separation energy spectra of two argon isotopes, which are compared to one-neutron removal experiments.
}
\maketitle
\section{Introduction}
\label{intro}
As ab initio calculations of atomic nuclei enter the A=40-100 mass range, a great challenge 
is to yield accurate predictions for the vast majority of open-shell isotopes. 
Together with the availability of nuclear Hamiltonians derived from chiral effective field theory (EFT) \cite{Epelbaum:2008ga} and the use of similarity renormalization group (SRG) \cite{Jurgenson:2009qs}, the remarkable development of many-body techniques for nuclear structure \cite{Cipollone:2013b, Roth2012, Hagen:2012fb, Hergert:2013uja, Holt:2012fr} is bringing us closer to this goal.

Self-consistent Green's function (SCGF) theory provides a powerful framework for the description of correlated nuclear systems \cite{Dickhoff:2004xx}. In the last few years, the method has been successfully applied to finite nuclei on the basis of realistic two-body Hamiltonians and for systems with a good closed-shell character \cite{Barbieri:2009ej, Barbieri:2009nx}. In these works the many-body expansion, in Green's function theory encoded in the truncation of the self-energy, incorporated particle-particle, hole-hole and particle-hole correlations in a consistent way \cite{Barbieri:2009ej}. Recently, efforts have been made to improve the SCGF approach in two directions. 
First, three-nucleon interactions have been implemented \cite{Cipollone:2013b}. Mandatory for a qualitative description of nuclei and nuclear matter, three-body forces represent one of the current frontiers in nuclear structure calculations, specially concerning the medium-mass region above $pf$-shell nuclei.
Second, the approach has been extended to the Gorkov formalism, which allows to address systems in which pairing correlations play a determinant role \cite{Soma:2011GkvI, Soma:2013rc}. This has opened a route to the ab initio description of full isotopic chains, enlarging the domain of applicability of SCGF by an order of magnitude. 
We report here on these recent advances and discuss selected results on nitrogen, oxygen, fluorine and argon isotopes.

\section{Self-consistent Green's function approach}

In Green's function theory the many-body Schr\"odinger equation is recast into the Dyson equation
\begin{equation}
\label{eq:Dyson}
G_{\alpha \beta}(\omega) = G_{\alpha \beta}^{\text{HF}}(\omega) + \sum_{\gamma \, \delta} \,
G_{\alpha \gamma}^{\text{HF}}(\omega) \, \Sigma^{\star}_{\gamma \delta}(\omega) \, G_{\delta \beta}(\omega) \, ,
\end{equation}
which is solved for the dressed propagator $G$ and where $\alpha, \beta, ....$ label a complete single-particle basis, $G^{\text{HF}}$ represents the Hartree-Fock propagator and $\Sigma^{\star}$ the irreducible self-energy.
In the present work we employ the algebraic diagrammatic construction to third order (ADC(3)) \cite{Schirmer1983, Barbieri:2007Atoms} for the self-energy, which reads
\begin{equation}
\label{eq:selfenergy}
\Sigma^{\star}_{\alpha \beta}(\omega) = \Sigma^{\infty}_{\alpha \beta} + 
\sum_{n \, n'} C_{\alpha n} \left [ \frac{1}{\omega-M} \right ]_{n n'} C^{\dagger}_{n' \beta} +
\sum_{p \, p'} D_{\alpha p} \left [ \frac{1}{\omega-N} \right ]_{p p'} D^{\dagger}_{p' \beta} \, , 
\end{equation}
where $M, N$ are interaction matrices in the $2h1p, 2p1h$ spaces and $C, D$ contain the couplings to single-particle states. $\Sigma^{\infty}$ represents the static self-energy.
The self-energy is constructed starting from the Hamiltonian $H(A) = H - T_{\text{c.o.m.}}(A) = U(A) + V(A) + W$, where we correct for the centre-of-mass kinetic energy and $U$, $V$ and $W$ collect all one-, two- and three-body contributions respectively. 
Three-body terms are included by means of one- and two-body effective interactions, derived after contractions of $W$ with correlated density matrices \cite{Cipollone:2013b}. 
% Such a procedure goes beyond standard normal ordering, which has been showed to account for most of the correlations associated to three-nucleon forces (3NF).
We use two- and three-body forces from chiral EFT at N$^3$LO and N$^2$LO respectively \cite{N3LO:2003, Navratil:2007zn} evolved to low momentum scales by free-space SRG \cite{Jurgenson:2009qs}.
Once the one-body propagator is computed, the ground-state energy can be obtained from a Koltun sum rule corrected for the presence of 3NF
\begin{equation}
E_0^A = \sum_{\alpha \beta} \frac{1}{4 \pi i} \int d \omega \left[ U_{\alpha \beta} + \omega \, \delta_{\alpha \beta} \right] \, G_{\beta \alpha} (\omega)
-\frac{1}{2} \, \langle \Psi_0^A \,  | \, W \, | \, \Psi_0^A \rangle \, .
\end{equation}
\begin{figure}[b!]
\centering
\vspace{-0.5cm}
\hspace{-0.2cm}
\includegraphics[trim=0 0 10 0,width=0.42\textwidth]{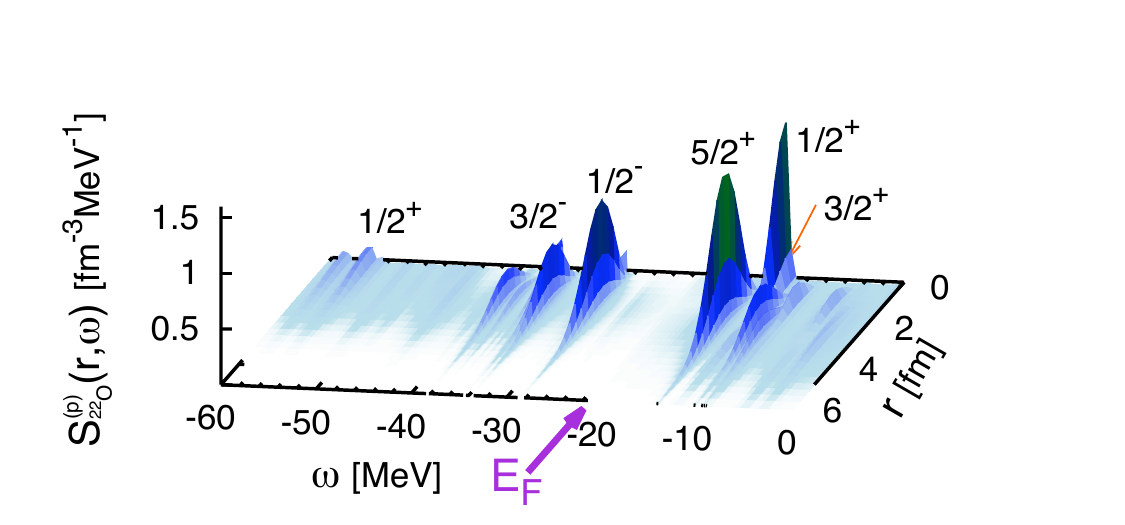}
\hspace{-0.0cm}
\includegraphics[trim=0 0 0 0,width=0.57\textwidth]{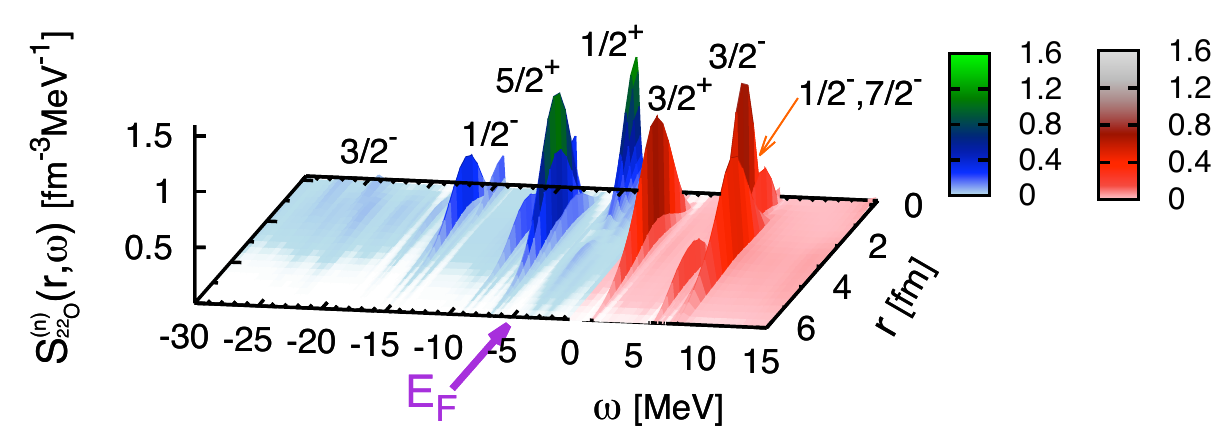}
\caption{ \textit{Left}: Diagonal part of proton spectral function in $^{22}$O. Strength on the left (right) of the Fermi energy $E_{\text{F}}$ corresponds to states in $^{21}$N ($^{23}$F). \textit{Right}: Diagonal part of neutron spectral function in $^{22}$O. Strength on the left (right) of the Fermi energy $E_{\text{F}}$ corresponds to states in $^{21}$O ($^{23}$O). Scattering states are shown in red.}
\label{fig:Ospectral}       
\end{figure}
%Focusing on selected oxygen isotopes, we first assessed the model-space convergence as well as the dependence on the renormalization scale $\lambda_{\text{SRG}}$, finding relative variations of about $1\%$ \cite{Cipollone:2013b}.

Within this scheme we have evaluated ground-state energies along the oxygen chain, which were found to be in excellent agreement with experiment and with other ab initio calculations \cite{Cipollone:2013b, Hergert:2013uja}.
In addition to yielding observables in the $A$-nucleon system, the propagator contains information on the separation energy spectra associated with the removal or addition of one particle. Such information is encoded in the spectral function, whose diagonal part (expressed in the coordinate basis) is
displayed in Fig. \ref{fig:Ospectral} 
for protons (left) and neutrons (right) in the case of $^{22}$O.
One sees how states close to the Fermi surface show a quasiparticle character, while strong fragmentation is present at large negative energies. The positive energy domain shown in the neutron spectral function corresponds to $^{23}$O scattering states.

\begin{figure}[t!]
\centering
\sidecaption
\includegraphics[width=8cm,clip]{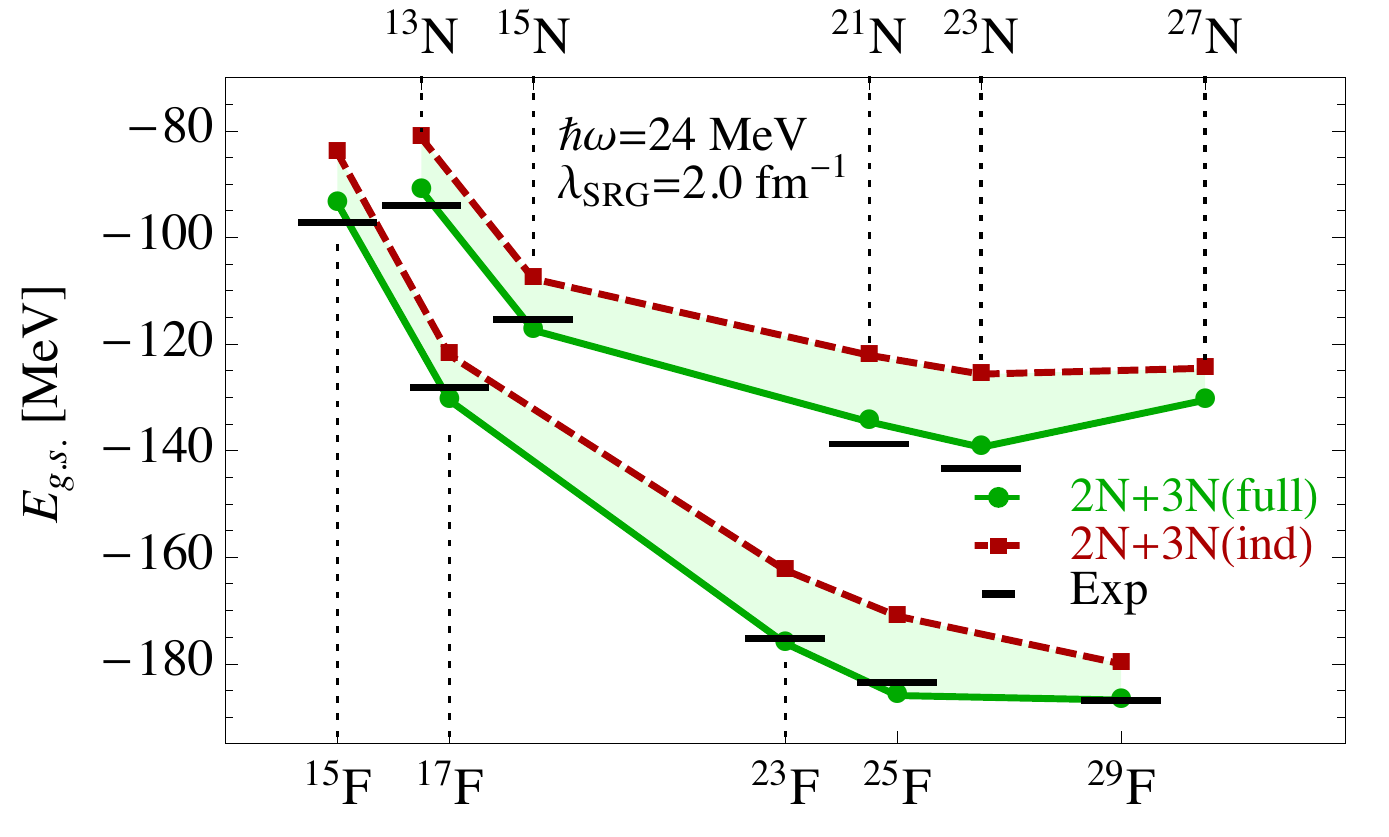}
\caption{Ground-state energies of nitrogen and fluorine isotopes. 
Calculations have been performed with chiral two- and three-body interactions evolved to $\lambda=2.0 \, \text{fm}^{-1}$ by means of SRG techniques.
SCGF results including induced only and full three-body forces are displayed and compared to experimental values.}
\vspace{-0.5cm}
\label{fig:NFenergies}       
\end{figure}

The peaks closest to $E_{\text{F}}$ in the spectra of Fig. \ref{fig:Ospectral} refer to ground states of $A\pm1$ nuclei. Using this information one can evaluate ground-state energies of odd-even systems in the  neighboring isotopic chains, namely nitrogen and fluorine. This can be done provided that corrections coming from the different c.o.m. kinetic energy in the $A$ and $A\pm1$ nuclei are properly taken into account \cite{Cipollone:2013b}.
In Fig. \ref{fig:NFenergies} we show the resulting binding energies of nitrogen and fluorine. The two curves correspond to calculations with only 3NF induced by the SRG procedure or with the full two plus three-body Hamiltonian that has been consistently SRG-evolved. 
Leading-order (N$^2$LO), original three-body terms are crucial in both bringing calculated energies close to experiment and yielding a correct description of the drip line.
The mechanism responsible for the drip line is the same as in the case of oxygen isotopes and
relates to the additional repulsion supplied by 3NF to the neutron $d_{3/2}$ orbit \cite{Cipollone:2013b}.

\section{Extension to open-shell nuclei}

Recently a SCGF scheme that can address open-shell nuclei has been introduced \cite{Soma:2011GkvI, Soma:2013rc}. The method is based on Gorkov's formalism \cite{Gorkov:1958}, which relies on the idea of trading the breaking of particle number with an effective description of pairing correlations. 
%which play a crucial role in open-shell systems.
%Particle number has to be eventually restored, which yields additional correlations in the systems but is not done at the present stage. 
Additional (anomalous) propagators that account for the breaking and formation of Cooper pairs are introduced. Formally, Eqs. \eqref{eq:Dyson} and \eqref{eq:selfenergy} still hold with all quantities (propagators and self-energies) being now matrices in a $2 \times 2$ Gorkov space. 
In order to work with the correct number of protons and neutrons in average, chemical potentials are included into the original Hamiltonian, i.e. we consider 
$\Omega(A) = H(A) - \mu_{\text{p}} \, Z - \mu_{\text{n}} \, N$.
\begin{figure}[b!]
\centering
\sidecaption
\includegraphics[width=4cm,clip]{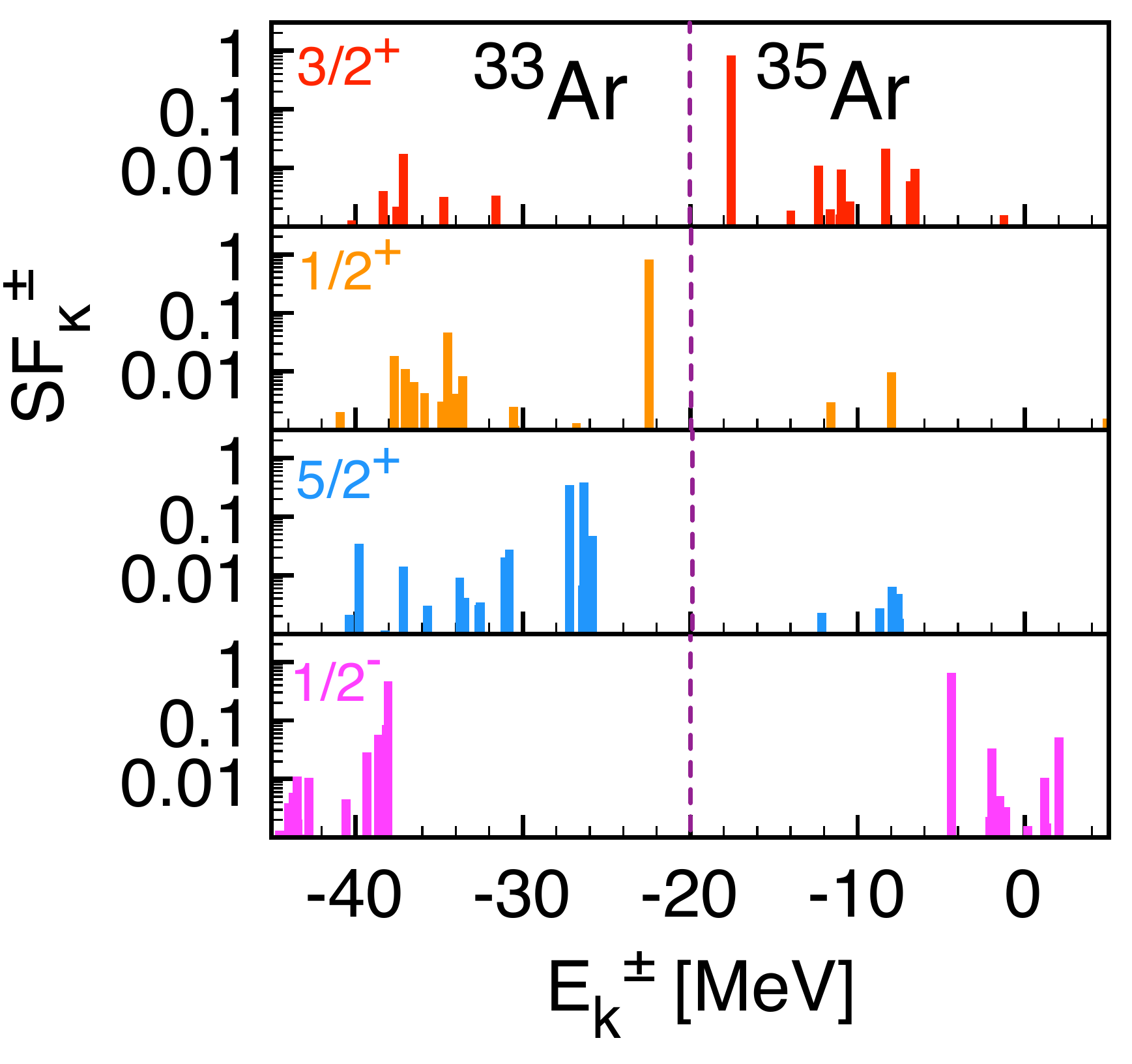}
\includegraphics[width=4cm,clip]{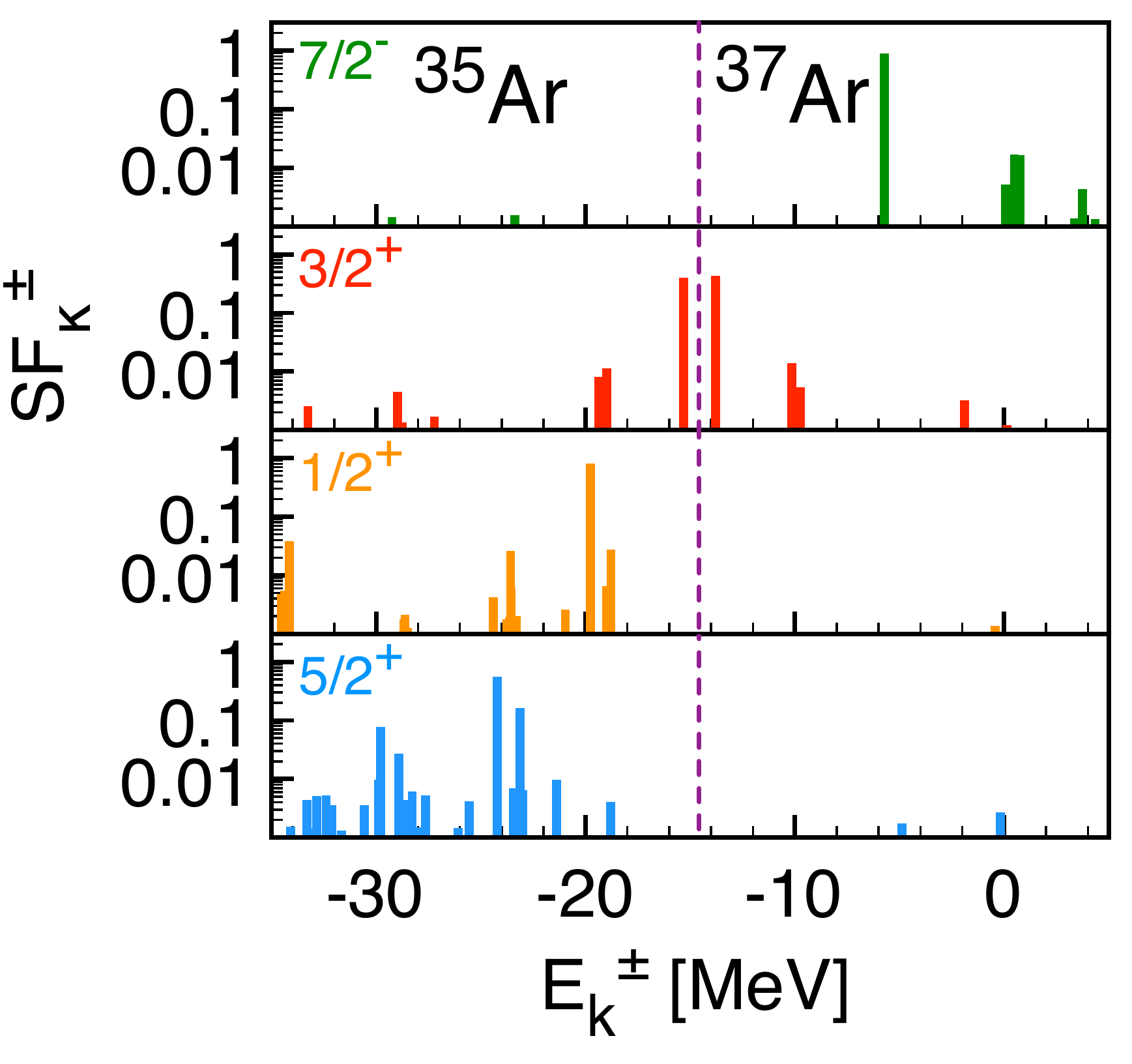}
\caption{Neutron spectral strength distributions in $^{34}$Ar and $^{36}$Ar from Gorkov-Green's function calculations. 
The dashed vertical line represents the (neutron) Fermi energy of the $A$ system $\mu_{\text{n}}$.
For different $J^{\Pi}$ states of the $A-1$ (left of $\mu_{\text{n}}$) and $A+1$ (right of $\mu_{\text{n}}$) nuclei, separation energies are shown with the associated spectroscopic factor.}
\label{fig:argon}      
\end{figure}

Within the Gorkov framework we have addressed the calculation of argon isotopes. While the input is the same two- plus three-body chiral interaction described above, the self-energy expansion differs, being limited to an ADC(2) scheme \cite{Soma:2011GkvI}. Extension to ADC(3) is currently in progress.
In Fig.~\ref{fig:argon} we display the spectral strength distributions for neutrons in $^{34}$Ar and $^{36}$Ar, obtained from the spectral functions by tracing over single-particle basis states.
From Fig. \ref{fig:argon} one can read out the one-neutron separation energy spectrum and the associated spectroscopic factors. In particular, the spectrum can be compared to data from neutron transfer reactions performed by Lee and coworkers at the National Superconducting Cyclotron Laboratory at Michigan State University~\cite{Lee:2010zzb}. The calculated one-neutron removal separation energies read $22.4 \, \text{MeV}$ and $15.3 \, \text{MeV}$  for $^{34}$Ar and $^{36}$Ar respectively, whereas the experimental values are $17.07 \, \text{MeV}$  and $15.25 \, \text{MeV}$. Other peaks in the spectrum can be compared to the ones observed experimentally, showing an overall qualitative agreement. 
The doubly open-shell $^{36}$Ar compares better with data than the semi-magic $^{34}$Ar, in particular in the vicinity of the Fermi surface. This suggests that pairing correlations are qualitatively accounted for. Coupling to collective fluctuations, absent in the present ADC(2) scheme, is expected to compress the calculated spectra and bring them closer to experiment.

\section{Conclusions}
SCGF calculations represent a valuable candidate for the ab initio description of atomic nuclei, providing information on ground-state properties as well as on excitation spectra of odd-even systems.
The inclusion of 3NF results crucial for a consistent description of nitrogen, oxygen and fluorine drip lines. Preliminary applications to argon demonstrate that the method, extended to a Gorkov scheme, can successfully address open-shell systems and aim at full medium-mass isotopic and isotonic chains.

\section*{Acknowledgments}

This work was supported by the DFG through grant SFB 634, the Helmholtz Alliance Program with contract HA216/EMMI, the UK Science and Technology Facilities Council under Grant 
ST/J000051/1 and Canada's NSERC (Grant No. 401945-2011). TRIUMF receives funding via a contribution through the Canadian National Research Council.
Calculations were performed using HPC resources from GENCI-CCRT through Grant 2013-050707 (France) and the DiRAC facility of STFC (UK).

 \bibliography{INPC13bib}

\end{document}